\documentstyle[12pt,aasms4]{article}

\def\etal{{\it et al.}}
\def\ie{{\it i.e.}}
\def\eg{{\it e.g.}}
\def\ta{{\cal A}}
\def\aveta{\overline{{\cal A}}}
\def\lap{\hbox{${_{\displaystyle<}\atop^{\displaystyle\sim}}$}}

\begin{document}

\title{STATISTICS OF GAMMA RAY BURST TEMPORAL ASYMMETRY}
\author{Bennett Link\altaffilmark{1}}
\affil{Montana State University, Department of Physics, Bozeman, MT 59717}
\altaffiltext{1}{Also Los Alamos National Laboratory}
\and
\author{Richard I. Epstein}
\affil{Los Alamos National Laboratory, Mail Stop D436, Los Alamos, NM 87545}

\begin{abstract}

We study the temporal asymmetry of over 600 bursts from the BATSE 3B
catalog, encompassing a 200-fold range in peak flux. By comparing the
rates of rise and fall of the flux near the highest burst peak, we
find that about two-thirds of the bursts exhibit a preferred asymmetry
in the sense that the flux rises more rapidly than it falls,
confirming the conclusions of previous studies employing smaller
databases. The statistical significance of the average time asymmetry
of the sample is $>99.999$\%; therefore, models that predict time
symmetry of the burst profile are ruled out.  We find no statistically
significant correlation between burst temporal asymmetry and peak. This
result is consistent with both cosmological and local interpretations
of the gamma ray burst phenomenon.

\end{abstract}

\section{INTRODUCTION}

The origin of gamma ray bursts (GRBs) remains a mystery since their
detection by the {\em Vela} satellites in July 1969
(\cite{discovery}).  The Burst and Transient Source Experiment (BATSE)
on board the {\em Compton Gamma Ray Observatory} continues to detect
GRBs distributed with striking isotropy on the sky, and with a dearth
of faint events compared to that expected for a homogeneous
distribution (\cite{isotropy1}; \cite{isotropy2}).  With counterparts
at other wavelengths yet to be identified, the distances to GRBs
remain uncertain. The idea that GRBs originate at cosmological
distances has emerged as a serious possibility, while the
prospect that bursts originate within the halo of the Galaxy remains
tenable. In the face of these uncertainties, the identification and
interpretation of fundamental properties of GRB variability is needed
to assess proposed models.

Cosmological GRB models account naturally for the observed isotropy
while explaining the dearth of faint events as due to a modification
of the observed distribution by the universal expansion.  These models
also predict a general time dilation of the more distant, fainter 
events. Recently, some authors have found indications of time
dilation consistent with cosmic expansion (see, \eg, \cite{dilation1};
\cite{dilation2}; \cite{dilation3}; \cite{dilation4}), but the evidence is
not yet statistically compelling.  Models in which GRBs originate near
the Galaxy have the appeal of modest luminosity requirements compared
to cosmological models, but need fine tuning to maintain
consistency with the observed isotropy and number-flux relationship.

Some GRB models, such as those accounting for burst time structure as
due soley to beams crossing our line of sight, predict time-symmetric
light curves. Recent quantitative studies of the shapes of GRB light
curves, however, have established that GRBs are time-asymmetric
(\cite{LEP}; \cite{norris_asymmetry}; \cite{timeasymmetry};
\cite{mitrofanov}; \cite{fishman_asymmetry}), confirming ealier claims
(see, \eg, \cite{barat_asymmetry}). Link, Epstein, \& Priedhorsky
(1993) utilized a {\em skewness function}, similar to that used by
Weisskopf \etal\ (1978), and found that the majority of 20 bright GRBs
selected from the first 48 detected by BATSE are time-asymmetric in
the sense that the flux rises more rapidly than it falls. In a
subsequent study, Nemiroff \etal\ (1994) quantified the degree of time
asymmetry by considering the ratio of the number of times where the
counts in a given time bin are lower than in the previous bin to the
number of the times the counts are higher.  Nemiroff \etal\ (1994)
studied about 40 bright bursts, and confirmed with high statistical
significance the result found by Link, Epstein, \& Priedhorsky (1993).
Mitrofanov \etal\ (1994) constructed an average light curve from 260
bursts that showed a quickly rising flux followed by a slower decay.
The results of these studies exclude models that predict
time-symmetric GRB light curves.

Inasmuch as temporal asymmetry is independent of burst duration and
intensity, it is a useful statistic to compare with other burst
properties.  The purpose of this paper is to study temporal asymmetry
for a sample of bursts encompassing a large range in peak flux, and to
test for a correlation between temporal asymmetry and  peak flux.
The discovery of a correlation would provide evidence for evolutionary
or cosmological effects, though the lack of such a correlation would
not necessarily constitute evidence against the cosmological
interpretation. We find that approximately two-thirds of
bursts spanning a 200-fold range in peak flux exhibit the same
temporal asymmetry, with no statistically significant correlation
between the asymmetry and the peak flux.

In \S 2 we describe the analysis techniques we use in studying burst
temporal asymmetry and the attributes of the data set tested. In \S 3
we discuss our results, and in \S 4 we summarize our results and their 
implications. 

\section{ANALYSIS}

As a measure of the shape of GRB light curves, we define a 
{\em time-asymmetry} parameter as the third-moment of the burst time profile:
\begin{equation}
\ta\equiv {\langle (t - \langle t \rangle)^3 \rangle
               \over
         \langle (t - \langle t \rangle)^2 \rangle^{3/2}}, 
\end{equation}
where $\langle\rangle$ denotes an average over the data
sample performed as: 
\begin{equation}
\langle f(t) \rangle \equiv {\sum_i (c_i - c_{th}) f (t_i)
               \over \sum_i (c_i - c_{th})}. 
\end{equation}
Here $c_i$ is the measured number of counts in the $i$th bin, $t_i$ is
the time of the $i$th bin, and $c_{th}$ is a threshold count level.
The time-asymmetry parameter is calculated for a contiguous data
sample including the burst peak and nearby bins in which the counts
exceed $c_{th}$. We define the threshold as 
\begin{equation}
c_{th} \equiv f (c_p - b) + b, 
\end{equation}
where $c_p$ is the peak count rate, $b$ is the background, and $f(<1)$
is a fraction that will be fixed for each data set. For a given $f$,
this definition of the threshold ensures that $\ta$ is calculated to
the same fraction of the peak flux, relative to the background, for
each burst in the data sample. Larger values of $f$ emphasize the
structure of the peak over that of the surrounding foothills. The
normalization of $\ta$ has been chosen in such a way as to make it
independent of burst amplitude, duration and background. For a
time-symmetric burst peak, $\ta=0$, and $\ta>0$ ($<0$) for a burst
whose flux rises (falls) more quickly than it falls (rises). For an
infinitely fast rise followed by an exponential decay, $\ta=2$,
independent of the time-constant of the decay.

\section{RESULTS}

In Fig. 1 we show the burst time-asymmetry parameter $\ta$ as a
function of peak flux for 631 bursts from the BATSE 3B catalog,
selected as described below.  Our sample contains faint bursts as well
as bright ones, spanning a 200-fold range in peak flux. We use the
BATSE PREB plus DISC data types at 64 ms time resolution, with the
four energy channels 25-50 keV, 50-100 keV, 100-300 keV, and $>300$
keV combined to attain the best statistics.  For each burst, $\ta$ was
calculated for a contiguous sample of data containing the highest
burst peak and for which each bin satisfies $c_i\ge c_{th}$. The only
requirement for a burst to be tested is that the data sample satisfying
$c_i\ge c_{th}$ contain at least three bins.

The error bars in Fig. 1 represent one-$\sigma$ deviations from the
calculated time asymmetry due to photon counting statistics; they were
obtained by randomizing the measured counts according to Poisson
statistics and calculating the variance of the time-asymmetry
parameter for many trials. A preponderance of positive time
asymmetries is apparent in Fig. 1 for all peak fluxes; about
two-thirds of the bursts have positive time asymmetry. In Table 1 we
show the weighted average of the time-asymmetry parameter for
different values of $f$. The small values of the standard deviation in
Table 1 show that the positive time asymmetry is not an artifact of
Poisson noise. To estimate the statistical significance of the
positive time asymmetry, we consider the probability that the observed
fraction of positive $\ta$ bursts occurs by chance. For example, for
our largest sample in Table 1 containing 631 bursts, 68\% of the
bursts have positive temporal asymmetry. If $\ta$ is a random variable
of zero mean, the probability of such a high percentage of $\ta>0$
bursts occurring by chance is $\lap 0.001$\%. 

While most bursts have positive time asymmetry, there are numerous
counterexamples. Examples of bursts with positive and negative $\ta$
are shown in Fig. 2. Quite often the $\ta>0$ events have a
simple structure of rapid rise followed by slow decline, while the
$\ta<0$ events have multiple peaks, as illustrated in the figure. This
result is consistent with the conclusion of Bhat \etal\ (1994) 
that bursts with more complex structure are less frequent.

In Fig. 1 it appears that time asymmetry increases with peak flux. Is
this apparent trend statistically significant? In Table 2 we show the
number-averages of $\ta$ computed for the bright and dim halves of the
sample. For our largest data set ($f=0.1$), $\ta_{\rm bright}/\ta_{\rm
dim}$ is $\sim 2$; for larger values of $f$, the ratio is larger.
However, because noise has zero average time asymmetry, we expect
fainter events, which have lower signal to noise than bright events,
to exhibit $\ta$ values closer to zero.  To estimate the significances
of the $\ta_{\rm bright}/\ta_{\rm dim}$ values found, we studied how
the $\ta$ values of the bright half of bursts change upon degrading
them to fainter peak fluxes. With each burst in the bright half of the
sample, we identified a peak flux selected at random from the dim
half.  Each bright burst was then degraded by reducing the counts in
each bin by a factor of the peak count rate of the selected dim burst
to the peak count rate of the bright burst, $c_{p,{\rm dim}}/c_{p,{\rm
bright}}$.  To each bin in this simulated dim burst, we added a
Poisson-deviate with a mean of the new number of counts. In this way
we produced simulated dim bursts with the same intrinsic temporal
asymmetry as the bright bursts. To estimate the significances of the
ratios $\ta_{\rm bright}/\ta_{\rm dim}$ in Table 2, we produced
numerous simulated data sets, calculated the ratio for each
simulation, and determined the frequency with which the simulated
$\ta_{\rm bright}/\ta_{\rm dim}$ exceeded the value from the real data set. The
simulated value of $\ta_{\rm bright}/\ta_{\rm dim}$
exceeded the value measured from the real data set in 39\% to 56\% of the
simulations, depending on $f$. We conclude that the apparent trend of
$\ta$ with brightness in the BATSE 3B catalog is not statistically
significant.\footnote{Computing weighted averages $\aveta$ in the bright
and dim halves of the data set also shows no statistically
significant dependence of $\ta$ on peak flux.}

\section{SUMMARY AND CONCLUSIONS}

We have applied a simple measure of burst temporal asymmetry to a
large, uniform sample of bright and dim bursts. About two-thirds of
GRBs have time-asymmetric peaks in the sense that the flux rises more
rapidly than it falls, confirming the results of previous analyses of
bright bursts (\cite{LEP}; \cite{timeasymmetry}; \cite{mitrofanov}).
We conservatively estimate the significance of the preferred time
asymmetry at over 99.999\%, thus excluding GRB models that predict
time-symmetric light curves. We find that the preferred time asymmetry
shows no significant dependence on peak flux. This result is
consistent with both cosmological and local interpretations of the GRB
phenomenon.

\acknowledgements

It is a pleasure to thank E. E. Fenimore and J. S. Bloom for providing
us with data from the BATSE 3B catalog in a form convenient for this
analysis. This work was performed under the auspices of the U.S.
Department of Energy, and was supported in part by NASA EPSCoR grant
\#291471.

\newpage

\figcaption{Burst time-asymmetry parameter $\ta$ for 631 GRBs. 
The threshold was chosen
by taking $f=0.1$. Error bars represent one-$\sigma$ deviations of
$\ta$ from the measured values. The vertical line divides the bright
half of the sample from the dim half.}

\figcaption{Examples of bursts of positive temporal asymmetry (a), and
negative temporal asymmetry (b). }

\newpage

\begin{deluxetable}{cccccc}
\tablewidth{0pt}
\tablecaption{Burst Time Asymmetry}
\tablehead{
\colhead{$f$} & \colhead{sample size} & \colhead{events with $\ta>0$} & 
\colhead{$\aveta$\tablenotemark{a}}
& \colhead{$\sigma$} & \colhead{confidence}}
\startdata
0.1 & 631 & 68\%  & 0.14 & 0.00041 & 99.9993 \nl    
0.2 & 603 & 66\%  & 0.092 & 0.00048 & 99.99  \nl 
0.5 & 463 & 62\%  & 0.086 & 0.0013 & 99.1 \nl
0.67& 350 & 59\%  & 0.12 & 0.0024 & 91 \nl
\tablenotetext{a}{Weighted averages, $\aveta\equiv\sum_i \ta_i
\sigma_i^{-2}/\sum_i \sigma_i^{-2}$, where the variance is $\sigma^2=(\sum_i\sigma_i^{-2})^{-1}$}
\enddata
\end{deluxetable}

\begin{deluxetable}{ccc}
\tablewidth{30pc}
\tablecaption{Average Time Asymmetry -- Bright vs. Dim Bursts}
\tablehead{
\colhead{$f$} & \colhead{$\ta_{\rm bright}/\ta_{\rm dim}$}\tablenotemark{a} & 
\colhead{likelihood}\tablenotemark{b}}
\startdata
0.1      &    1.7      &    56\% \nl
0.2      &    2.0      &    39\% \nl
0.5      &    2.7      &    $\sim 50$\% \nl
\footnotesize
\tablenotetext{a}{Number averages are calculated by giving equal weights to
all data points, \ie, $\ta_{\rm bright}\equiv N^{-1}\sum_{\rm
bright}\ta_i$.}
\tablenotetext{b}{Likelihood that the quoted $\ta_{\rm
bright}/\ta_{\rm dim}$ is spurious, based on the percentage of 
simulations yielding values of 
$\ta_{\rm bright}/\ta_{\rm dim}$ greater than the quoted value.} 
\enddata
\end{deluxetable}

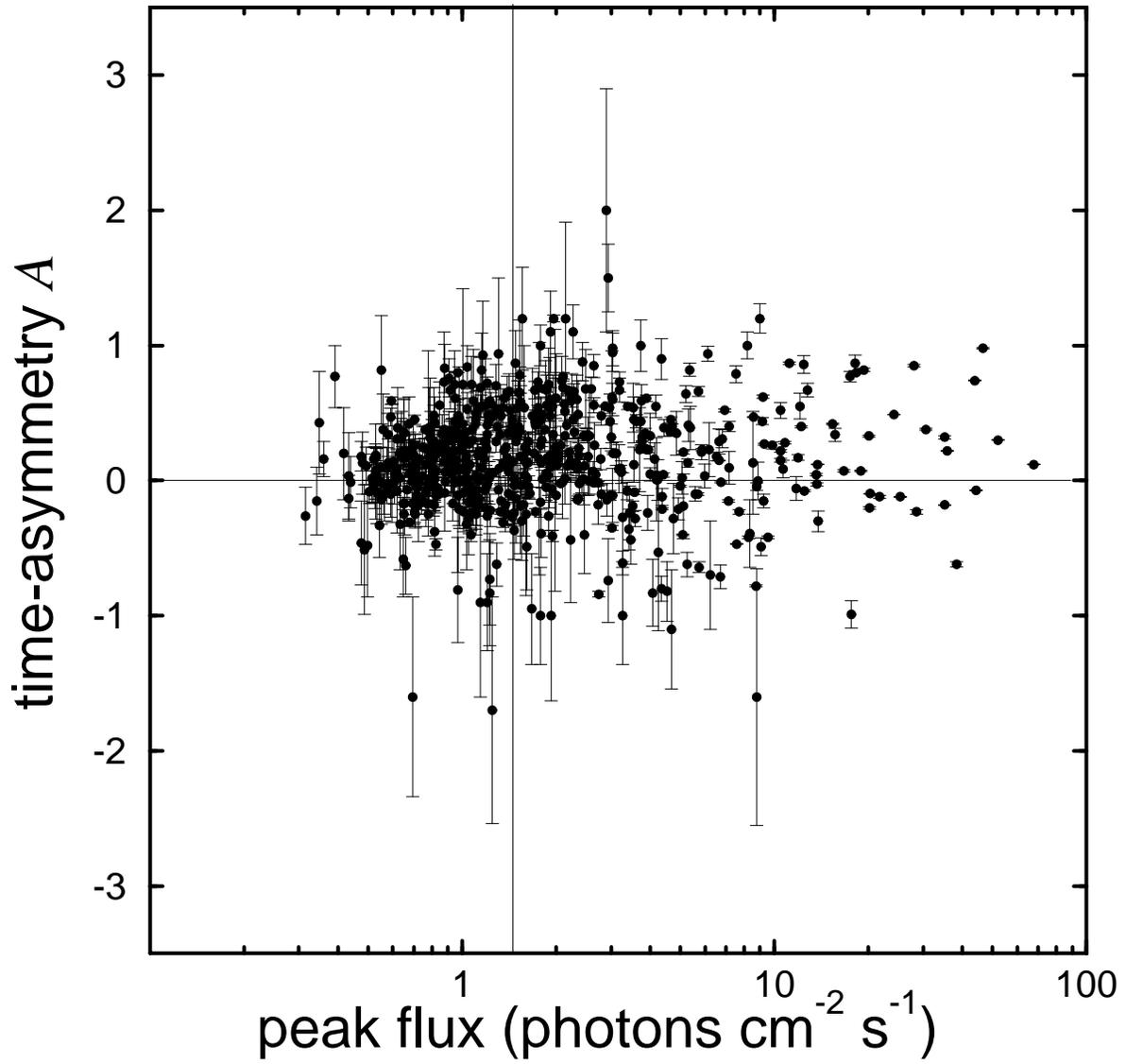
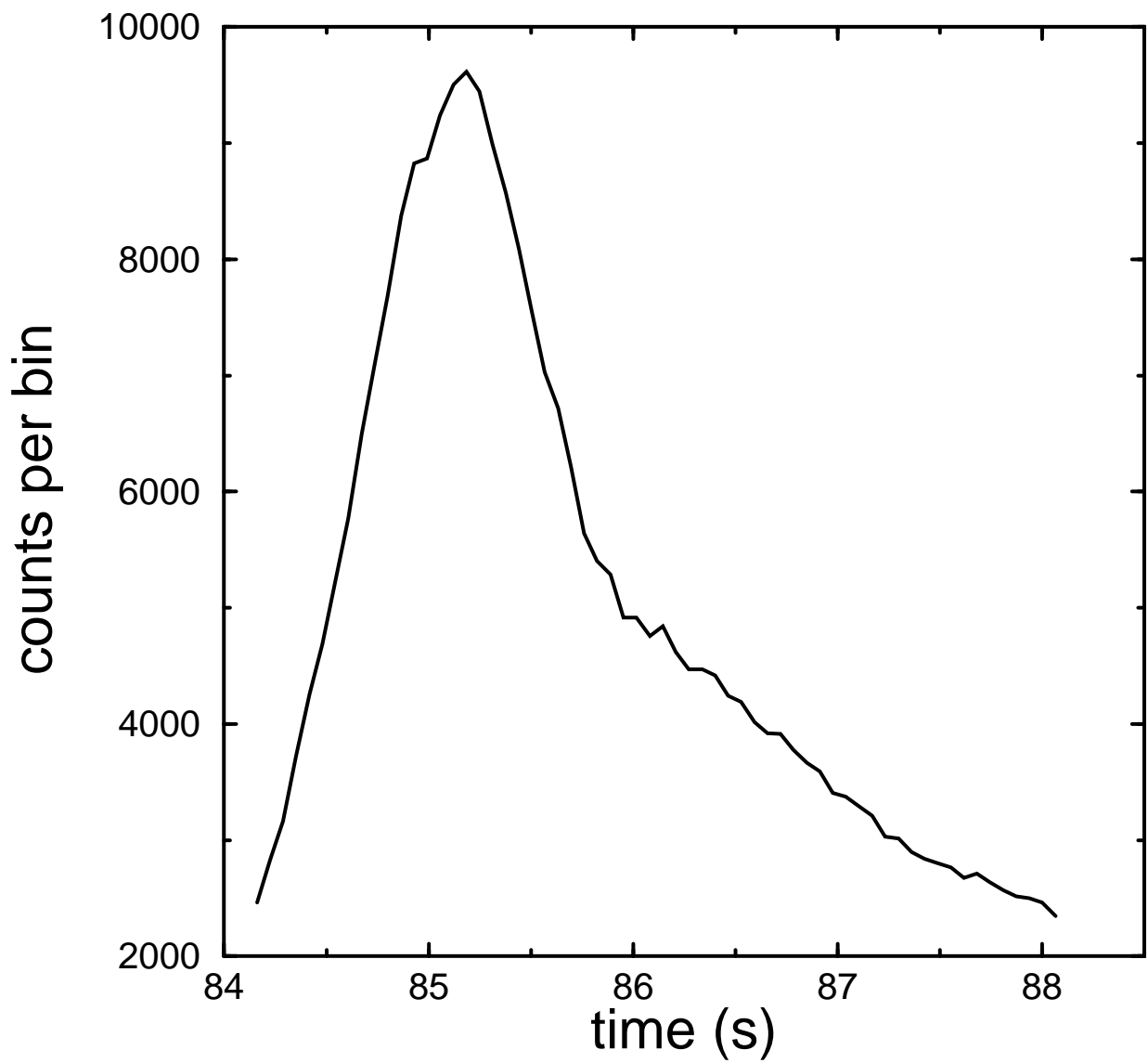
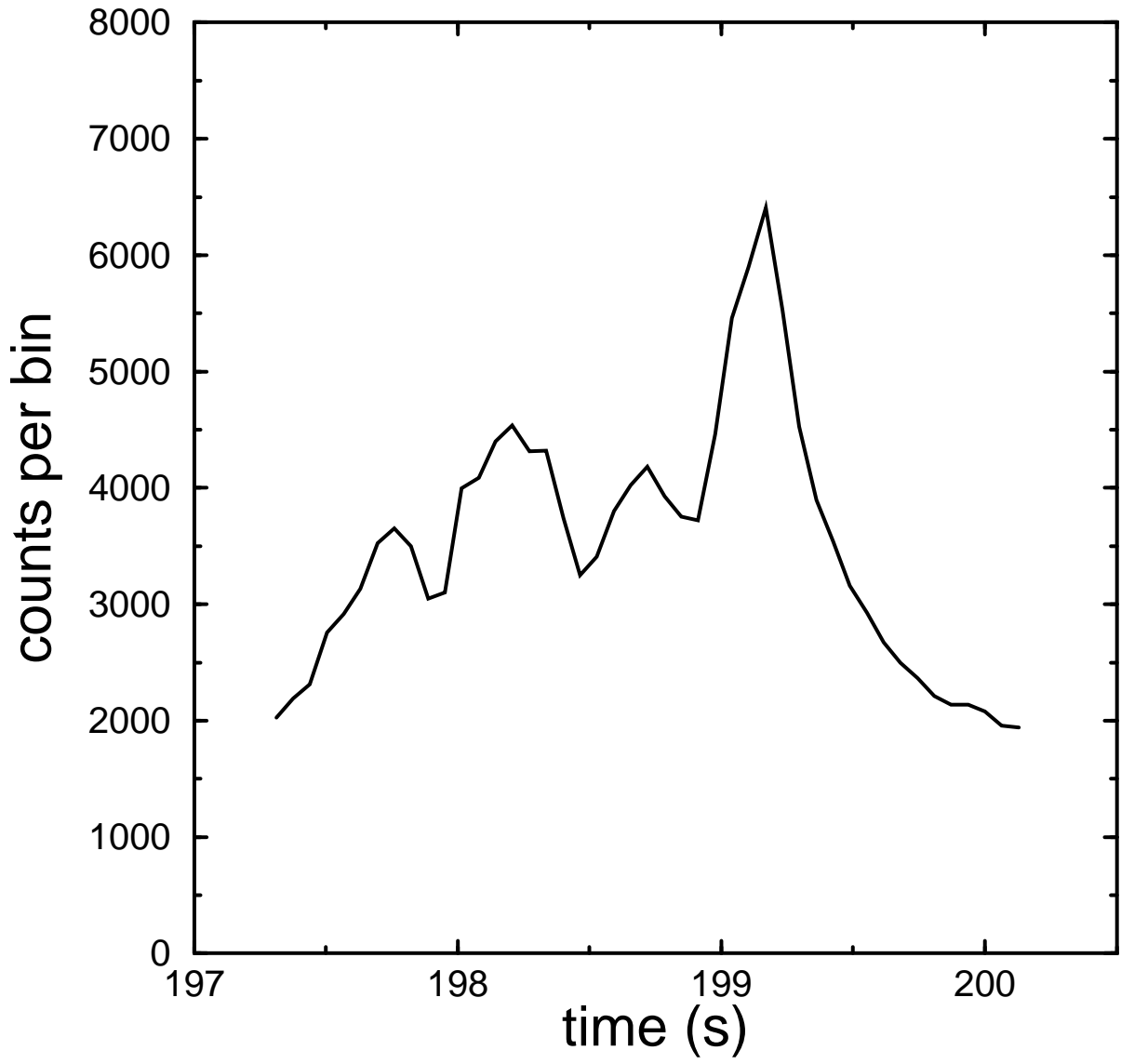
\end{document}